# Iron-based trinuclear metal-organic nanostructures on a surface with local charge accumulation

Cornelius Krull[1], Marina Castelli[1], Prokop Hapala[2*], Dhaneesh Kumar[1], Pavel Jelinek[2,3] and Agustin Schiffrin[1,4*]

[1]*School of Physics & Astronomy, Monash University, Clayton 3800 Australia*
[2]*Institute of Physics of the CAS, Prague 16200 Czech Republic*
[3]*RCPTM, Palacky University, Šlechtitelů 27, 783 71, Olomouc, Czech Republic.*
[4]*ARC Centre of Excellence in Future Low-Energy Electronics Technologies Monash University, Clayton 3800 Australia*
*Corresponding author: agustin.schiffrin@monash.edu*
*Corresponding author: hapala@fzu.cz*

Coordination chemistry relies on harnessing active metal sites within organic matrices. Polynuclear complexes – consisting of organic ligands binding to clusters of several metal atoms – are of particular interest, owing to their electronic/magnetic properties and potential for functional reactivity pathways. However, their synthesis remains challenging; only a limited number of geometries and configurations have been achieved. Here, we synthesise – via supramolecular chemistry on a noble metal surface – one-dimensional metal-organic nanostructures composed of terpyridine (*tpy*)-based molecules coordinated with well-defined polynuclear iron clusters. By a combination of low-temperature scanning probe microscopy techniques and density functional theory, we demonstrate that the coordination motif consists of coplanar *tpy*'s linked via a linear tri-iron node in a mixed (positive) valence, metal-metal bond configuration. This unusual linkage is stabilized by a local accumulation of electrons at the interface between cations, ligand and surface. The latter, enabled by the bottom-up on-surface synthesis, hints at a chemically active metal centre, and opens the door to the engineering of nanomaterials with novel catalytic and magnetic functionalities.

Metal-organic molecular complexes and coordination polymers allow for a vast range of functionalities, both technological and biological, from catalysis to light-harvesting and gas storage, sensing and exchange.[1-3] In these processes, the atomic-scale electronic configuration of the metal centres plays a crucial role, since it dictates the chemical reactivity of the molecular systems.[1-3] For instance, polypyridyl-based complexes, due to their coordination morphology and resulting opto-electronic and chemical properties, can be used for photovoltaics[4] and catalysis.[5] Polynuclear complexes are of special interest, since magnetic[6] and electronic interactions between metal atoms in close proximity can give rise to useful properties (e.g., cooperative electronic and steric effects), with multiple active metal centres potentially enhancing catalytic processes in comparison to mononuclear systems.[7],[8,9] Wet chemistry synthesis methods have achieved an array of different polynuclear complexes, exhibiting direct metal-to-metal bonds.[10,11] Based both on inorganic compounds and organic matrices stabilising transition metal centres, a variety of geometries have been obtained, e.g., trigonal nodes, linear nanochains up to 8 atoms long.[10] However, the synthesis of such compounds remains challenging, with only a limited number of configurations available, and metals and ligands utilised. Notably, only few[10] metal-to-metal compounds based on iron (an important element for catalysis[12]) exist.

On-surface supramolecular chemistry[13] – a bottom-up approach driven by noncovalent intermolecular interactions and metal-ligand bonding on a surface – allows for the design of self-assembled, atomically precise metal-organic nanomaterials, with morphologies, properties and functionalities that can differ dramatically from those obtained via conventional synthetic chemistry. Such approaches can provide an alternative for synthesising functional complexes with polynuclear metal sites.[13] Confining the molecular and atomic building units to two-dimensions (2D) can lead to the stabilisation of morphologies/functionalities not achievable in three-dimensions (3D), and can be beneficial for device-based applications, e.g., solid-state heterogeneous catalysis,[14] where the active materials are required to interface with a solid. The surface, via its structural symmetry and chemical reactivity, can further provide a means of control over atomic-scale morphology and electronic/chemical properties.[13] These methods have recently allowed for the synthesis of low-dimensional metal-organic systems with di-nuclear coordination metal centres (e.g., di-iron complexes with promising catalytic [15] and magnetic[16] properties). Higher-order polynuclear systems have also been achieved, e.g., tri- and tetra-copper coordination.[15,17] However, multinuclear complexes with direct metal-to-metal interactions (and their resulting properties)[8] have not been rigorously demonstrated on surfaces. A trinuclear macromolecular complex based on the coordination of a terpyridine (*tpy*)-containing molecule and Cu adatoms from a Cu surface has been proposed,[17] but without a direct atomic-scale characterisation of the coordination motif. Here, we report the bottom-up on-surface synthesis and atomic-scale characterisation of a tri-nuclear Fe complex, resulting from the one-dimensional (1D) self-assembly of a *tpy*-based aromatic molecule (*tpy-phenyl-phenyl-tpy;* TPPT; chemical structure in inset of Fig. 1b) coordinated with Fe adatoms on a Ag (111) surface. Using a combination of low-temperature scanning tunnelling microscopy (STM) and spectroscopy (STS), non-contact atomic-force-microscopy (ncAFM), local-contact-potential-difference (LCPD) measurements, density-functional-theory (DFT) and ncAFM image simulations, we directly demonstrate that the metal-ligand coordination motif consists of flat, coplanar *tpy*'s linked via a linear tri-iron cluster. The Fe atoms in this coordination node are in a cationic mixed valence configuration, exhibiting direct metal bonding. Importantly, the coordination is concomitant with a charge redistribution, with electrons accumulating at the cation-ligand-surface interface, stabilizing the metal-organic linkage, and potentially leading to catalytic activity localized at the multinuclear centre.

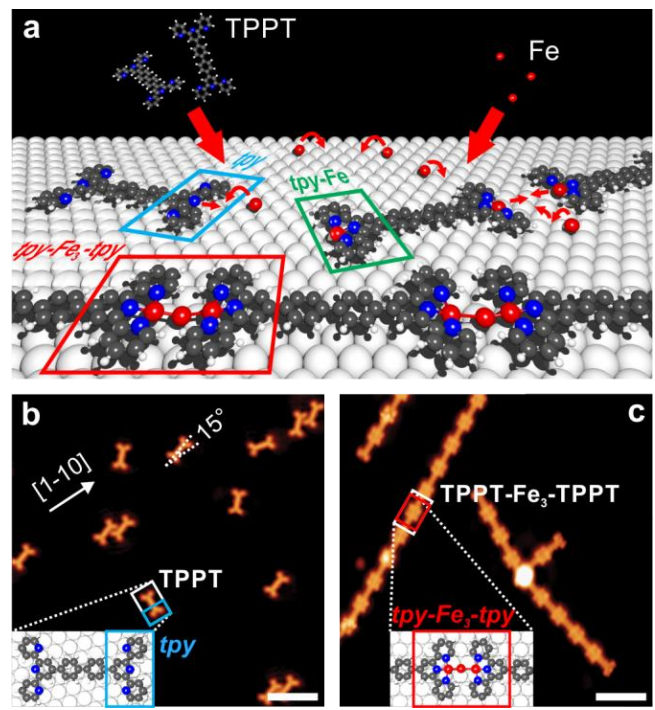

**Figure 1. Self-assembly of 1D metal-organic nanostructures on a noble metal surface. a,** Schematic of the bottom-up synthesis of metal-organic nanochains (MOCs) via on-surface terpyridine (tpy)-iron (Fe) coordination. TPPT-molecules and Fe adatoms were sequentially deposited from the gas-phase onto the Ag(111) surface held at room temperature. **b,** STM image of TPPT molecules adsorbed on Ag(111) ($I_t$ = 5 pA, $V_b$ = 20 mV). **c,** STM image of MOCs resulting from the on-surface metal-ligand coordination ($I_t$ = 25 pA, $V_b$ = 20mV). The STM tip was functionalised by picking up a single carbon monoxide molecule (Methods). Insets: DFT-calculated energetically favourable adsorption geometries of TPPT (a) and MOC (b; black: carbon; blue: nitrogen; red: iron; grey: silver). Scale bars: 5 nm.

Figure 1b shows an STM image of TPPT molecules on Ag(111) (see Methods for sample preparation). As other aromatic molecules on noble metals,[18] TPPT adsorbs flat and is imaged as a "dog-bone" feature with "v-shaped" *tpy*'s, resembling its Lewis structure (see Ref. [19] for more details). Subsequent deposition of Fe results in the formation of self-assembled 1D nanostructures (Fig. 1c). We observed metal-organic-chains (MOCs) with lengths of several tens of molecular units, that is, on the order of 100 nm. These MOCs consist of TPPT-molecules with their *tpy*'s arranged in a head-to-head motif, stabilized by the coordination with Fe adatoms.

In Figures 2a-c, we imaged TPPT with STM in different stages of coordination: single pristine TPPT; TPPT with one of its *tpy*'s coordinated with an Fe adatom; two TPPT's linked by Fe within a MOC. The non-metalated left *tpy* of the molecule in Fig. 2b is imaged identically to that of the pristine molecule (Fig. 2a), with its characteristic "v-shape". The right metalated *tpy* in Fig. 2b appears brighter, with a central protrusion, due to the interaction with Fe. The metal-organic linkage in the MOC consists of apparently flat *tpy*'s in a symmetric head-to-head configuration, with a central protrusion (Fig. 2c). STM does not resolve the intramolecular conformation, nor the atomic-scale configuration of the metal-organic node mediating the chaining.

To elucidate the atomic-scale morphology of the 1D nanostructures, we performed ncAFM with CO-functionalised tips [(Supplementary Information (SI)]. This technique allows for real-space imaging with sub-molecular resolution.[20] Figures 2d-f show CO-tip ncAFM images of the same systems as in Figs. 2a-c. The image of pristine TPPT (Fig. 2d) is similar to the Lewis-structure of the gas-phase molecule, confirming quasi-flat adsorption. The two phenyl (*ph*) rings are slightly tilted out of the molecular plane (SI). The pyridine (*pyr*) of the *tpy*'s appear as distorted hexagons,[21] tilted towards the surface due to interactions between the nitrogen lone electron pairs and the noble metal.[19] NcAFM-imaging of the singly metalated TPPT (Fig. 2e) shows a left non-metalated *tpy* identical to that of the pristine molecule (Fig. 2d), consistent with STM; the central *ph*'s remain tilted. Imaging of the right *tpy* (coordinated with one Fe adatom) differs from the left, non-metalated one. The distal *pyr*'s are tilted, with the lower side oriented towards the centre of the *tpy*, indicating a rotation around the C-C bond to enable coordination between the N atoms (Fig. 2d white arrows) and Fe. This coordination-induced conformational change is expected.[22] The in-plane angle γ between the distal *pyr* is reduced by the metalation (Fig. 2d; SI). The axial *pyr* appears darker in comparison to that of the non-metalated *tpy*, with its N-termination barely visible. This indicates an increased attraction between the CO-tip and the *tpy-Fe,* and a bending of this *pyr* due to bonding between Fe and surface (Fig. 2e profiles), consistent with previous work on molecules interacting with metal adatoms.[23] The ncAFM map in Fig. 2f shows a coordination node in a MOC, with two opposing *tpy*'s and a central bright feature. We estimate[21] a distance of 8.4 ± 1 Å between opposing axial N-atoms, more than twice as large as that of the analogous, single metal complex in solution.[24] Each *tpy* within the node is similar to the *tpy-Fe* in Fig. 2e: the darker sides of the distal *pyr*'s are rotated towards the centre of the *tpy*; the axial *pyr* shows the same darker depression due to increased tip-surface attractive force (SI). We deduce that the coordination node is composed of at least two Fe atoms, with each of the two *tpy*'s interacting with one adatom (as for *tpy-Fe* in Figs. 2b, e).

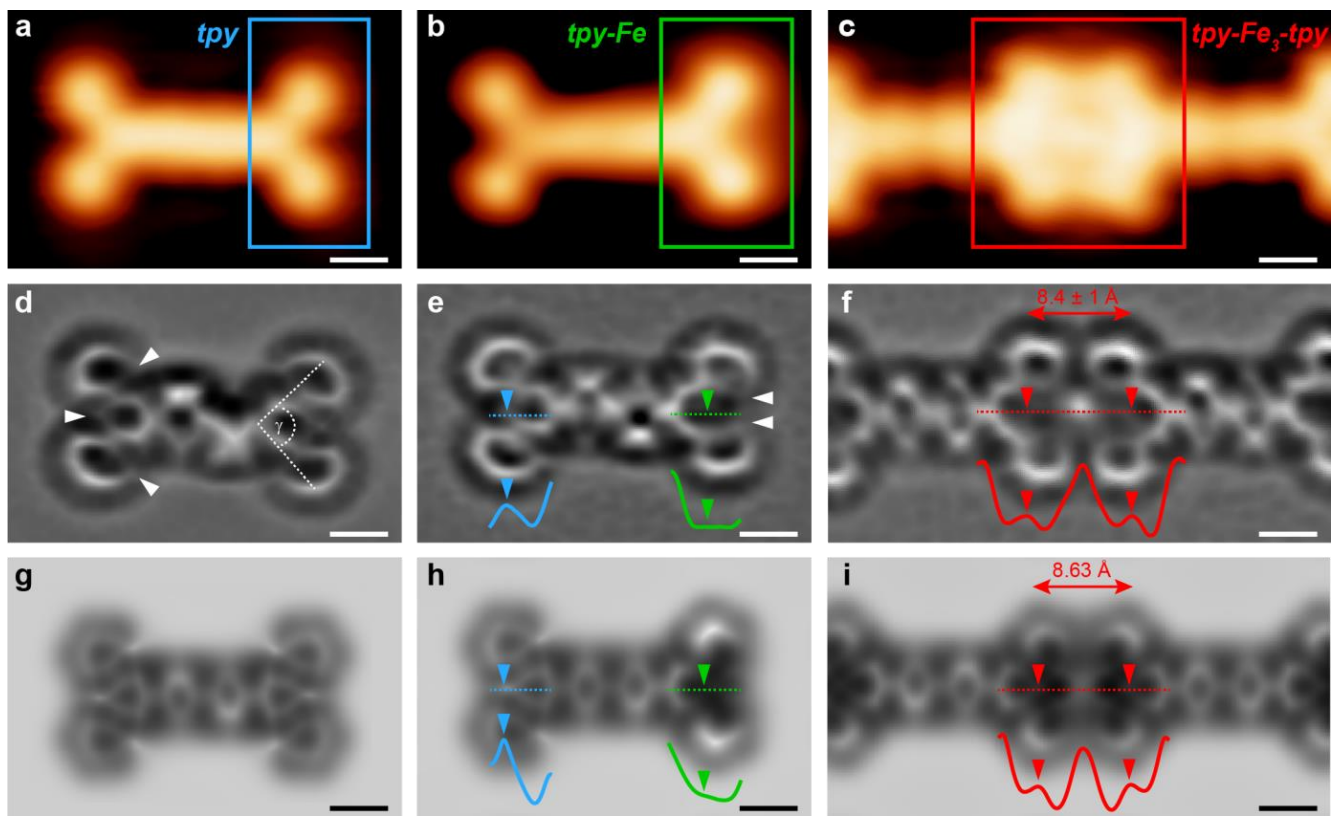

**Figure 2. Intramolecular morphology of isolated TPPT molecule (a, d, g), singly metalated TPPT (b, e, h) and metal-organic coordination node in chain (c, f, i) on Ag(111). a-c**, STM images with CO-functionalised tip (a: $I_t$ = 5 pA, $V_b$ = 20 mV; b, c: $I_t$ = 10 pA, $V_b$ = 20 mV). **d-f**, Constant-height Laplace-filtered ncAFM images with CO-functionalised tip [tip height is 0.8 Å above that determined by the STM setpoint $I_t$ = 25 pA, $V_b$ = 20 mV measured on bare Ag(111)]. **g-i**, Simulated ncAFM images based on DFT calculations (Methods). Coordination node in (i) includes 3 linearly arranged Fe atoms. Locations of N atoms correspond to lower intensity areas of *pyr* [indicated by white arrows in (d) and (e)]. Inset curves: apparent height profiles [arb. u.] along the dashed lines. Scale bars: 5 Å.

To determine the atomic-scale morphology of the metal-organic linkage, we *deconstructed* a Fe-TPPT node in a MOC via STM lateral manipulation (Figs. 3a, b), revealing its composition: a *tpy* imaged identically to a *tpy* metalated with a single Fe adatom (Fig. 3c), labelled *tpy-Fe$^{(D)}$* (green box); and a *tpy* imaged with an elongated protrusion (orange box, *tpy-Fe$_2^{(D)}$*). Analogously, by manipulating a single Fe adatom towards a *tpy* coordinated with a single Fe adatom (*tpy-Fe$^{(A)}$*), we *assembled* a *tpy* (*tpy-Fe$_2^{(A)}$*; Fig. 3e) imaged identically to *tpy-Fe$_2^{(D)}$*. This provides evidence that *tpy-Fe$_2^{(D)}$* consists of a *tpy* coordinated to two Fe adatoms, similar to di-nuclear poly-*pyr* coordination chains on Cu(111).[25] We did not observe spontaneous self-assembly of this doubly metalated *tpy-Fe$_2$* complex.

We performed d$I$/d$V$ STS (Fig. 3f-i) to corroborate that the *deconstructed* species (*tpy-Fe$^{(D)}$*, *tpy-Fe$_2^{(D)}$*) are identical to the respective *assembled* ones (*tpy-Fe$^{(A)}$*, *tpy-Fe$_2^{(A)}$*). The (d$I$/d$V$)/($I$/$V$) spectra in Fig. 3h (*tpy-Fe$^{(A)}$*) show tunnelling resonances at ~ +1.56 V (associated with an empty orbital at the centre[19] of TPPT) and > +1.8 V (related to the metalated *tpy*). This spectroscopic signature is identical to that of *tpy-Fe$^{(D)}$* (Fig. 3f). The spectra in Fig. 3i (*tpy-Fe$_2^{(A)}$*) show resonances at ~ +1.67 V (centre of molecule) and > +1.5 V (metalated *tpy*), equivalent to those of *tpy-Fe$_2^{(D)}$* (Fig. 3g). Our data provide compelling evidence that the metal-organic node consists of a singly metalated *tpy-Fe* group and a doubly metalated *tpy-Fe$_2$*, that is, a cluster of 3 linearly arranged Fe atoms.

We calculated the energetically favourable configurations of the molecular systems of interest via DFT, and simulated ncAFM images (Figs. 2g-i; Methods; SI). The latter reproduce our experiments: (i) the outwards pointing N atoms of the tilted distal *pyr*'s of the pristine TPPT (Fig. 2g); (ii) the coordination-mediated rotation of the distal *pyr*'s, with the N atoms oriented towards the centre of the dark, singly metalated *tpy-Fe* (Fig. 2h); (iii) the linear tri-iron coordination node with the central bright feature, indicating a repulsive electrostatic interaction between CO-tip and node (SI). Remarkably, the experimental and simulated apparent height line profiles (curves in Fig. 2d, e, h, i) are in excellent agreement, for a range of tip-sample distances (SI).

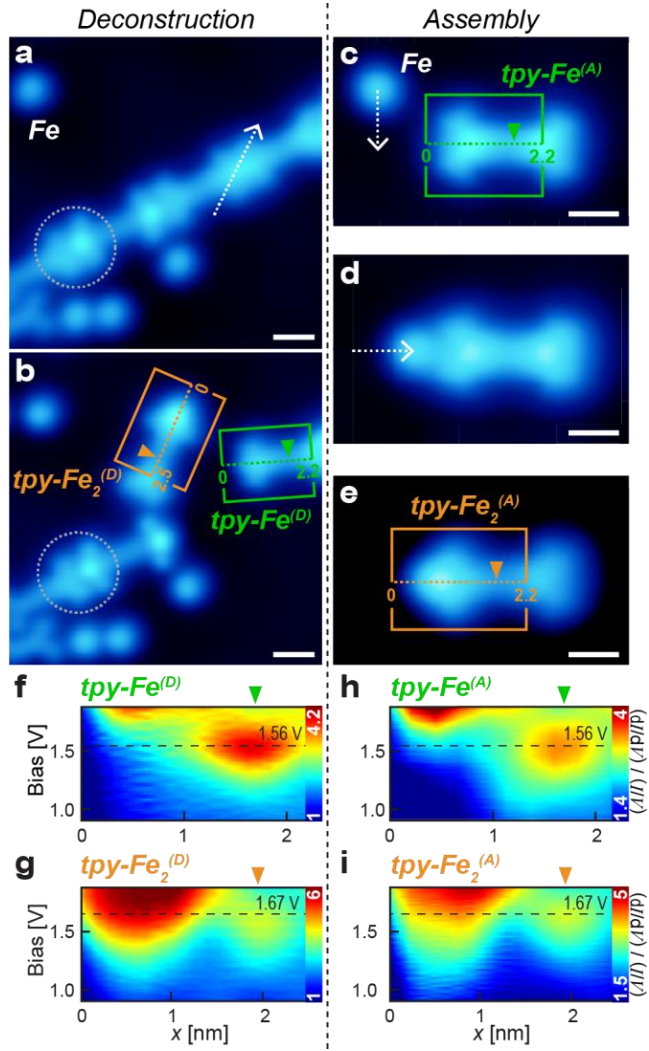

**Figure 3. Direct demonstration of tri-iron coordination node via STM manipulation. a-b,** STM images ($V_b$ = -500 mV; a: $I_t$ = 50 pA; b: $I_t$ = 400 pA) before and after lateral tip movement (white dashed arrow) across coordination node in MOC. The node is composed of two types of metalated *tpy*'s: *tpy-Fe$_2$*$^{(D)}$ and *tpy-Fe*$^{(D)}$. Grey dashed circle indicates coordination node decorated with a CO molecule (SI). **c-e,** STM images ($I_t$ = 100 pA, $V_b$ = -500 mV) of the assembly of a doubly metalated *tpy* (*tpy-Fe$_2$*$^{(A)}$) by manipulating a single Fe atom towards a singly metalated *tpy* (*tpy-Fe*$^{(A)}$). Imaging of singly metalated *tpy* (*tpy-Fe*$^{(D)}$, *tpy-Fe*$^{(A)}$) is different from Fig. 2b due to tunnelling parameters and no CO-tip functionalisation. **f-i,** (d$I$/d$V$)/($I$/$V$) spectra acquired along singly and doubly metalated *tpy*'s (orange and green dashed lines in b, c, e). Singly (doubly) metalated *tpy* shows tunnelling resonance at 1.56 V (1.67 V, respectively) associated with unoccupied state at centre of TPPT. Setpoint: $I_t$ = 25 pA, $V_b$ = -1 V. Topographic and electronic properties of *tpy-Fe*$^{(D)}$ (*tpy-Fe$_2$*$^{(D)}$) are identical to those of *tpy-Fe*$^{(A)}$ (*tpy-Fe$_2$*$^{(A)}$, respectively). STM manipulation and imaging performed with Ag-terminated Pt/Ir tip. Scale bars: 1 nm.

To understand the central protrusion in the ncAFM image in Fig. 2f, we performed local-contact-potential-difference (LCPD) mapping. This method can provide insight into the intramolecular charge distribution of the system.[26,27] In a MOC (Fig. 4a), the LCPD values $V^*$ show significant contrast between the TPPT ligand, Fe centre and Ag substrate. This is indicative of a negatively charged TPPT due to substrate-to-molecule electron transfer, similar to other aromatic systems.[18] Importantly, the map in Fig. 4b shows a maximum $V^*$ at the centre of the node, consistent with an accumulation of negative charge.

Figure 4e shows the LCPD along a metalated *tpy-Fe* (green curve) and a coordination node (red). The LCPD of the ligand within a MOC is larger (11 ± 6 mV) than for the isolated metalated TPPT. This is consistent with the larger energy of the (d$I$/d$V$)/($I$/$V$) resonance associated with an empty molecular state of *tpy-Fe$_2$* in comparison to *tpy-Fe* (Figs. 3h-i). This larger energy – indicating a smaller electron affinity for *tpy-Fe$_2$* – could be explained by a larger negative charge at the ligand due to further metalation of *tpy*. The LCPD at the Fe of the isolated metalated TPPT (green curve, Fig. 4e) does not show the increase observed for the centre of the coordination node (red curve); this is a signature of the tri-iron coordination, and hints towards a nontrivial charge redistribution at the node, not observed for the analogous single-Fe atom organometallic complex in solution.[22]

To explore this, we considered the DFT-calculated charge density at the node and the resulting electron density difference (Methods); see Fig. 4f. The latter shows electron depletion around the Fe cores (blue), and electron accumulation (red) between the Fe, N and Ag atoms, indicative of Fe-N coordination bond formation. Notably, the electrons between the Fe atoms point towards the formation of a metal-metal bond, with Fe-Fe distances (~2.4 ± 0.2 Å) comparable to other Fe-based multinuclear systems.[10] Moreover, significant negative charge from the Ag substrate contributes to stabilising the node.

The electron density involved in the Fe-Fe bonds and at the node-Ag interface (Fig. 4f) gives rise to a repulsive electrostatic interaction with the (negatively charged) CO molecule on the tip, that is, a negative electrostatic potential above the node (Fig. 4c). This is responsible for the increase of LCPD (Fig. 4b) and for the protrusion at the centre of the tri-iron linkage see in ncAFM imaging (Fig. 2f, i; SI).[28] These features are due to the node bonding motif and resulting electronic properties and electrostatic field, rather than to topography (the central Fe lies ~0.55 Å closer to the surface than the adjacent ones; Fig. 4d). Although the node is electrostatically negative, Bader analysis (see SI) yields a positive charge state for all three Fe atoms. Notably, the distal Fe (Bader charge ~ +0.8e) differs significantly from the central (~ +0.15e); the cluster has mixed valence, and the chemical state and reactivity of the cations – *central Fe less positive than distal* – depend on their location. The negative charge accumulation results from over-screening of the cations, mainly by the Fe-Fe bond electrons, with some contribution from the surface due to the central Fe bridging to the conduction bath.

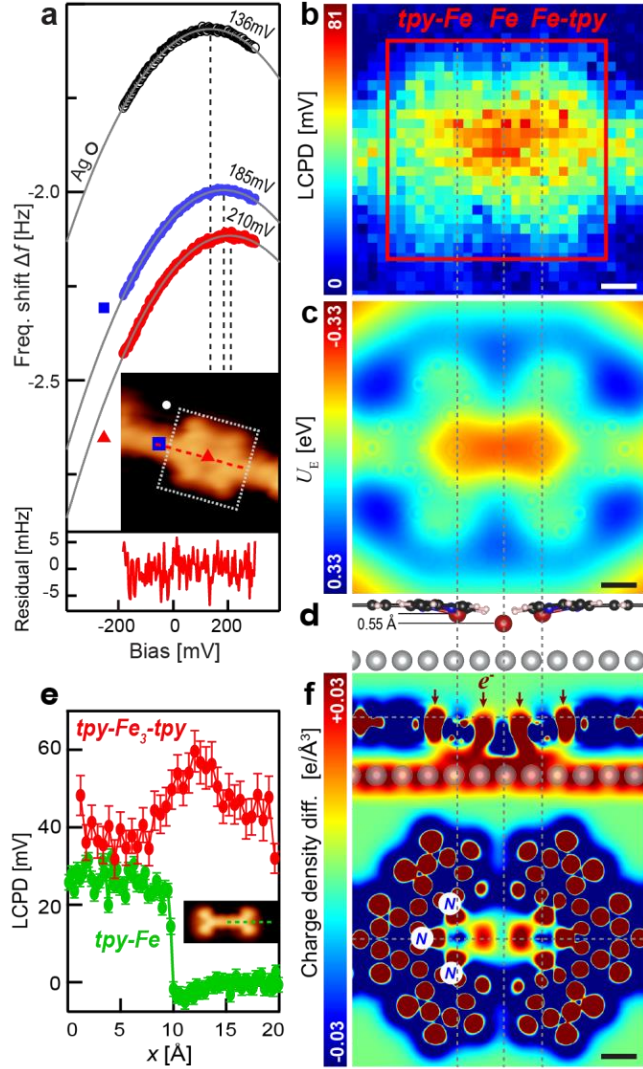

**Figure 4**. **Charge redistribution and electron accumulation at centre of tri-iron coordination node. a,** Frequency shift $\Delta f$ as a function of sample bias voltage $V_b$, measured at the centre of tri-iron node (red), on TPPT molecule in MOC (blue) and on bare Ag(111) (white). Data recorded at constant height, 3 Å above that defined by STM setpoint ($I_t$ = 25 pA, $V_b$ = 20 mV) on Ag(111). Value of $V_b$ for which $|\Delta f|$ is minimum corresponds to the LPCD (Methods). Grey curves: parabolic fitting of $\Delta f(V_b)$. Uniform fit residual indicates that tip-sample forces are mainly due to electrostatic interactions.[27] Inset: STM image of coordination node in MOC with CO functionalised tip ($I_t$ = 25 pA, $V_b$ = 20 mV). **b,** LCPD map of *tpy-Fe₃-tpy* node relative to Ag(111), indicating negative charge accumulation at the centre of node. **c,** DFT-simulated electrostatic potential at the node, 3 Å above molecular plane. **d,** DFT-calculated energetically favourable geometry of tri-iron node (lateral view). **e,** LCPD measured across tri-iron node [red; see dashed line in inset in (a)] and singly metalated *tpy* (green; dashed line in CO-tip STM image of Fe-TPPT-Fe in inset). Data recorded at constant height of 3 Å above that defined by STM setpoint ($I_t$ = 25 pA, $V_b$ = 20 mV) on Ag(111). **f,** DFT-calculated charge density difference at the coordination node (top: vertical cut; bottom: horizontal; dashed horizontal lines indicate cut positions), showing electron accumulation (vertical red arrows) between cations, ligand and surface. White circles indicate positions of N atoms of left *tpy* of node. Dashed vertical lines indicate Fe atom positions. Frequency shift and LCPD measurements performed with CO-functionalised Pt/Ir tip. Scale bars: 2 Å.

These charge distribution properties and electrostatics would differ drastically if the node contained two Fe's instead of three (SI). π

In conclusion, we presented the bottom-up synthesis of a tri-iron metal-organic coordination complex on a metallic surface. Importantly, our experiments – supported by DFT and ncAFM imaging simulation – directly demonstrate the mixed (positive) valence trinuclear configuration of the coordination motif, as well as the local electron accumulation at the ligand-cation-surface interface. Our on-surface approach is crucial for the formation of this unusual trinuclear complex. The noble metal surface offers balanced molecule-surface interactions, confining the molecular and atomic precursors to 2D and enabling efficient adsorbate diffusion. This results in a planar coordination motif, dramatically different from 3D, single-cation complexes in solution.[22] The coordination is stabilized by the site-specific charge redistribution at the metal-organic node. This mixed valence configuration can lead to specific on-surface sites that potentially exhibit local catalytic activity.[29,30] The self-assembly protocol can further allow for the design of well-defined 2D arrays of active multinuclear sites,[30] with nanoscale precision, providing an attractive platform for solid-state catalysis. Further applications could also benefit from the magnetic properties of metal-metal interactions in these nodes.[6]

**Methods**

**Sample preparation.** The 1D metal-organic nanostructures were synthesised in ultrahigh vacuum (UHV) by sequential deposition of TPPT-molecules (first) and iron atoms (second) from the gas-phase onto a clean Ag(111) surface (Mateck GmbH; prepared in UHV by repeated cycles of Ar⁺ sputtering and annealing at 720 K). See Fig. 1a for schematic. Terpyridine-phenyl-phenyl-terpyridine (TPPT; HetCat Switzerland) molecules were sublimed at 550 K onto the clean Ag(111) substrate held at room temperature (RT), resulting in a deposition rate of ~ $4 \times 10^{-4}$ molecules/(nm² s). Iron (Fe) atoms were subsequently deposited from the gas-phase with the substrate held at RT to allow for the metalation of TPPT molecules and the self-assembly of Fe-TPPT chains. The ratio of pristine to metalated molecules was controlled by varying the TPPT-to-Fe stoichiometry. The quantity and length of the chains can be controlled by the overall amount of metalated molecules on the surface as well as by the annealing time at RT (the longer the RT annealing time is, the longer the metal-organic chains). Both molecular and atomic adsorbates were deposited at sub-monolayer coverages. STM lateral manipulation experiments (Fig. 3) required further deposition of Fe with the substrate at 7 K to allow for single Fe adatoms on the surface. The base pressure was below $2 \times 10^{-9}$ mbar during molecular deposition, and below $1 \times 10^{-10}$ mbar during Fe evaporation and for all characterisation measurements.

**Carbon monoxide tip functionalisation**. Carbon monoxide (CO) molecules were used to functionalise both the STM and ncAFM Pt/Ir tips as follows.[31,32] CO was dosed *in situ* with the sample maintained below 8 K. By exposing the sample to a partial pressure $p(CO) = 5 \times 10^{-8}$ mbar, we reproducibly deposited CO at a rate of ~ 1 molecule/(1000 nm$^2 \cdot$s). Typical deposition times were 3 seconds. The tip was first positioned on top of a CO molecule, at a height defined by the STM setpoint ($I_t$ = 25 pA, $V_b$ = 20 mV), and then brought 400 pm closer to the surface to pick up the molecule. Once the CO was transferred from surface to tip, we confirmed that the molecule was adsorbed symmetrically by imaging another CO on the surface as a circular protrusion surrounded by a symmetric depression.[33]

**STM/STS measurements**. All STM and STS measurements were performed at 4.6 K with a Ag-terminated (blue colour mapping; Fig. 3) or a CO functionalised Pt/Ir tip for increased resolution (brown colour mapping; Figs. 1, 2a-c, insets Fig. 4). All topographic images were acquired in constant-current mode. STS measurements were obtained by measuring the tunnelling current as a function of tip-sample bias, scanning the region of interest pixel-by-pixel, with the tip height stabilized according to an STM setpoint at each location. The normalized numerical derivative (d$I$/d$V$)/($I$/$V$) spectra were computed as an approximation of the local density of states.[34] The sample bias is reported throughout the text.

**ncAFM measurements**. All ncAFM experiments were performed using a qPlus tuning fork sensor[35] (Createc) in frequency modulation mode at 4.6 K. (resonance frequency $f$ ~ 29 kHz; spring constant $k$~1800 N/m $\pm$ 7%), with a CO-terminated Pt/Ir tip. All ncAFM topography images were taken using a 0.6 Å amplitude at constant height. Various tip-sample distances were used, which were defined with respect to an STM setpoint on Ag(111) [$I_t$ = 25 pA, $V_b$ = 20 mV]. No bias voltage was applied during the ncAFM topographic measurements. All ncAFM topographic images were filtered by Laplace edge detection; SI for details.

**LCPD measurements.** LCPD measurements were performed with a qPlus tuning fork sensor (Createc) with an oscillation amplitude of 1 Å. All data were recorded at a constant height with respect to an STM setpoint on Ag(111). The LCPD $V^*$ – defined as the sample bias voltage $V_b$ for which the absolute value $|\Delta f(V_b = V^*)|$ of the resonance frequency shift is minimum – was determined by measuring $\Delta f$ as a function of $V_b$ (applied to the sample). The LPCD is plotted relative to that of the Ag substrate. In order to be sensitive to tip-sample forces which are dominantly electrostatic in nature, we performed LCPD measurements at *significantly large* tip-sample distances, for which $\Delta f$ ($V_b$) curves could be fitted with a parabola,[36] with a minimal random noise residual[27] (Fig. 4a). In these conditions, the LPCD $V^*$ is a measure of the variations of the electrostatic potential – and hence of the variations in charge distribution – at the surface. If we consider two different locations $\mathbf{r}_A$ and $\mathbf{r}_B$ on the surface, and $V^*(\mathbf{r}_A)$ and $V^*(\mathbf{r}_B)$ are positive (with respect to a zero-reference potential, in this case Ag), then there is an accumulation of negative charge at $\mathbf{r}_A$ and $\mathbf{r}_B$, and if $|V^*(\mathbf{r}_A)| > |V^*(\mathbf{r}_B)|$, then that charge accumulation is larger at $\mathbf{r}_A$ than at $\mathbf{r}_B$. The Pt/Ir tip was functionalised with a CO molecule for all ncAFM/LCPD measurements.

**STM lateral manipulation**. All STM lateral manipulation[37] experiments (Fig. 3) were performed at 4.6 K with an Ag-terminated Pt/Ir tip. The tip was approached onto a bare patch of Ag(111) by changing the STM setpoint ($I_t$ = 30 pA, $V_b$ = -10 mV for displacing Fe adatoms; $I_t$ = 100 nA, $V_b$ = -10 mV for breaking the coordination node), and subsequently moved laterally at a rate of ~500 Å/s.

**DFT calculations**. All DFT calculations were performed using the Vienna *Ab initio* Simulation Package (VASP)[38,39] with the projector augmented plane-wave method[40] and PBE exchange correlation functional. The empirical Grimme D2[41] correction with default parameters was used to model van der Waals interactions. The energy relaxation of the tri-iron metal-organic node was performed using [42] L(S)DA+U with two different values for $U$ ($U$ = 3.0 and 5.0 eV) to check the robustness of the results. Both calculations yielded the same symmetric node geometry, with Fe-Fe bond lengths of ~ 2.4 Å (in agreement with experiments) and antiferromagnetic ordering of Fe magnetic moments. It is important to note that calculations that did not include magnetic interactions or that included a ferromagnetic configuration led to asymmetric node geometries with significantly shorter Fe-Fe bond lengths, in disagreement with our experiments. The simulation supercell consisted of a rectangular Ag(111) slab of 17 $\times$ 8 atoms, containing two units of the molecular chain (total size: 49 $\times$ 20 $\times$ 20 Å$^3$). Initial geometric relaxations were performed with one fixed layer of Ag atoms. Simulations of ncAFM images (Fig. 2), electronic potential and charge density difference (Fig. 4) were obtained by considering 3 Ag layers and relaxing the electronic degrees of freedom while keeping the atom positions fixed. Charge densities and electrostatic potential did not exhibit any notable changes after adding Ag layers. The charge density difference corresponds to the subtraction between the charge density of the interacting system and that of the system composed of non-interacting atoms in the same positions as the relaxed system. Bader charge analysis was based on Ref. [43].

**ncAFM imaging simulations**. Simulated ncAFM images were obtained using the ProbeParticle code[44] with default van der Waals parameters (SI). The electrostatic potential was modelled as the electrostatic interaction between the DFT-relaxed system (see above) and the quadrupole model density on the probe particle (quadrupole strength Q = -0.4; stiffness of CO molecule: 0.5 N/m). Discrepancies between simulated and experimental images can be explained by: (i) overestimation of the delocalisation of $\pi$ electrons by DFT, resulting in flatter aromatic systems;[45] (ii) omission of long-range background forces in simulations, leading to a low frequency shift on the substrate; (iii) underestimation of the interactions between free electron pairs of N atoms of pristine TPPT with the metal surface, leading to bright protrusions in Fig. 2g.

**Acknowledgements**

M.C. and D.K. acknowledge support from the Monash Centre of Atomically Thin Materials (MCATM). A.S. acknowledges support from the Australian Research Council (ARC Future Fellowship FT150100426). P.J. acknowledges support from Praemium Academie of the Czech Academy of Sciences, the Ministry of Education of the Czech Republic Grant LM2015087 and GACR project No. 18-09914S. P. H. acknowledges support from Czech Academy of Sciences project MSM100101705.

# Iron-based trinuclear metal-organic nanostructures on a surface with local charge accumulation

*Supporting Information*

Cornelius Krull[1], Marina Castelli[1], Prokop Hapala[2*], Dhaneesh Kumar[1], Pavel Jelinek[2,3] and Agustin Schiffrin[1,4*]

[1]*School of Physics & Astronomy, Monash University, Clayton 3800 Australia*
[2]*Institute of Physics of the CAS, Prague 16200 Czech Republic*
[3]*RCPTM, Palacky University, Šlechtitelů 27, 783 71, Olomouc, Czech Republic.*
[4]*ARC Centre of Excellence in Future Low-Energy Electronics Technologies Monash University, Clayton 3800 Australia*
*Corresponding author: agustin.schiffrin@monash.edu
*Corresponding author: hapala@fzu.cz

# Contents

**Interaction between carbon monoxide and metal-organic complex**

Metal-organic complexes can be reactive to small molecules such as carbon monoxide (CO).[1] CO molecules introduced in the STM chamber – for CO-tip STM and ncAFM imaging (see main text) – can hence interact with the self-assembled nanostructures studied, even at extremely low coverages. To address this, we closely monitored the appearance of any new species that could potentially be related to interactions with CO, both for pristine and metalated TPPT. Figure S1 shows an STM topography image of pristine TPPT on Ag(111), before and after dosing CO. After dosing, small depressions appear on the Ag (fuchsia ticks in Fig. S1b). These features are due to individual isolated CO molecules.[2]

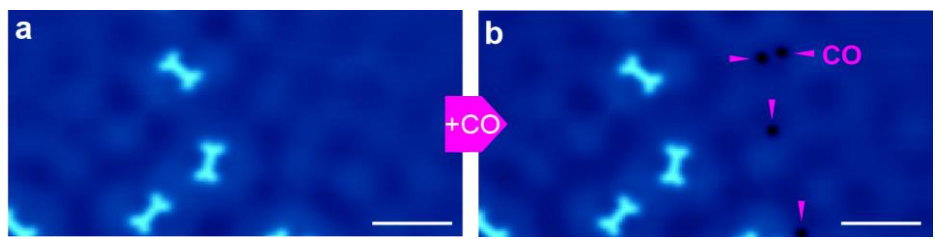

**Figure S1**: Constant-current STM topography image of TPPT/Ag111 ($I_t$ = 25 pA, $V_b$ = 20 mV) with a Pt/Ir tip, before (a) and after (b) dosing of carbon monoxide (CO). CO molecules are imaged as depressions, indicated with fuchsia arrows. Scale bar is 10 nm.

For Fe metalated TPPT molecules, we observed a new species that only appears after dosing CO. Figure S2a shows an STM image (Pt/Ir tip) of a TPPT molecule with each of its *tpy* groups coordinated with an Fe atom, after dosage of CO. Each *tpy-Fe* group is imaged as three protrusions, as opposed to the typical "hammer" shape of a non-decorated *tpy-Fe* (Fig. 3 in main text). By applying a bias voltage pulse close to one of the affected *tpy-Fe* groups in Fig. S1a (white cross), we could convert it to the "hammer"-shaped *tpy-Fe* (Fig. S2b). Moreover, the STM image after the pulse shows a new depression adjacent to the TPPT. To confirm that this depression is a CO molecule, we used our pick-up procedure to functionalise the STM tip (Methods), and subsequently imaged the same molecule. The resulting image in Fig. S2c shows the hallmarks of a CO-functionalised STM tip (Fig. 2 of main text). We thus conclude that the STM imaging of Fe-TPPT-Fe in Fig. S2a is the result of *tpy-Fe* groups interacting and binding with CO molecules.

Correspondingly, we also investigated interactions between CO and coordination nodes in the metal-organic nanochains (MOCs). Figure S2d shows an STM topography map of a MOC node (circled in grey), imaged differently (asymmetric) than the node discussed in the main text (symmetric). By lateral STM manipulation (Methods), we *deconstructed* this type of node. The STM imaging in Fig. S2e reveals that it is composed of a pristine *tpy-Fe* group and a CO-decorated *tpy-Fe* (Figs. S2e, f; compare with S2a, b). We unequivocally identify these species by identical STM imaging, for a series of different bias voltages ($V_b$ = -500 mV to 500 mV, data not shown). This type of asymmetric node is hence composed of two *tpy-Fe* groups bridged by a CO ligand. Our ability to identify CO-decorated species allows for the systematic investigation of pristine TPPTs, metalated TPPTs and MOCs. The results reported in the main text correspond to the latter, that is, systems that do not contain CO molecules.

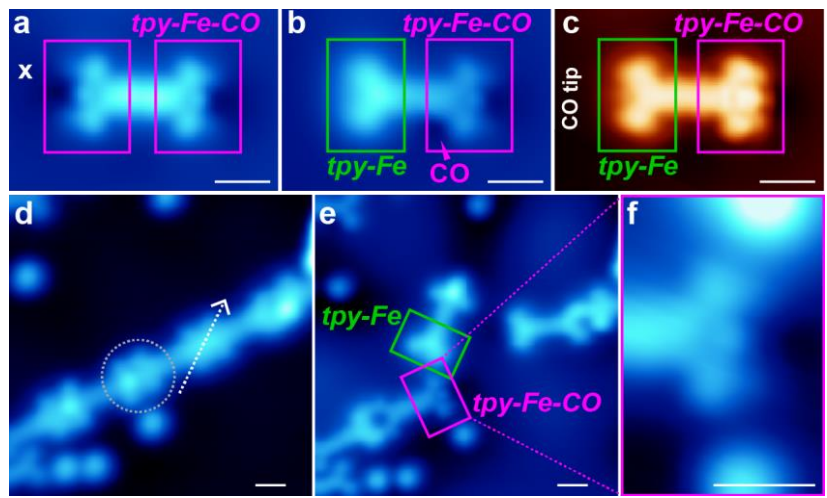

**Figure S2. Interaction between CO and Fe-TPPT complexes. a,** STM image of TPPT molecule with both *tpy-Fe* groups decorated with CO ($I_t$ = 50 pA, $V_b$ = -25 mV). **b,** STM image of the same molecule after applying a bias voltage pulse ($V_{pulse}$= 2 V, $t_{pulse}$= 200 ms, $I_t$ = 50 pA, $V_b$ = -25 mV) at the white cross: the CO molecule *jumped* next to one of the distal *pyr* rings, resulting in a pristine left *tpy-Fe* ($I_t$ = 50 pA, $V_b$ = -25 mV). **c,** STM image of molecule in (b), after picking up the adjacent CO molecule with the tip ($I_t$ = 50 pA, $V_b$ = -25 mV). **d,** STM image of MOC. Dashed white arrow indicates tip displacement direction during lateral STM manipulation across symmetric, non-decorated ($I_t$ = 400 pA, $V_b$ = -500 mV). **e,** STM image of *deconstructed* MOC: the node circled in (d) consists of a CO-decorated *tpy-Fe-CO* and a *tpy-Fe* ($I_t$ = 400 pA, $V_b$ = -20 mV). **f,** Zoomed-in image of *tpy-Fe-CO* in (e)**.** STM topography images were recorded with a Pt/Ir tip in (a), (b), (d)-(f) (blue colour coding) and with a CO-functionalised tip in (c) (brown colour coding). Scale bars are 1 nm.

**ncAFM experiments**

*Influence of tip-sample distance*

Non-contact AFM imaging (Fig. 2 of main text) was performed at a constant height of 0.8 Å above a reference height defined by the STM tunnelling setpoint on bare Ag(111): $I_t$ = 25 pA, $V_b$ = 20 mV. Using $I(z)$ measurements we estimated an absolute tip height of 6 ± 1 Å above the sample.

In addition, we performed ncAFM imaging for varying tip heights ranging from 1.4 to 0.6 Å above the same STM setpoint $I_t$ = 25 pA, $V_b$ = 20 mV (Fig. S3). As the tip approaches the molecule, short range repulsive forces become stronger and contribute significantly to the imaging, increasing the intramolecular chemical bond contrast (Fig. S3m).[3,4] At small distances (0.4 Å; 0.6 Å; 0.8 Å), the *pyr* rings of the molecule exhibit a repulsive feature at their centre, which has been observed for other aromatic molecules and is attributed to bending of the CO molecule on the tip.[5] At 0.4 Å above the setpoint (Fig. S3k, l), both the *tpy* and *tpy-Fe* groups display these repulsive features. However, their appearance differs significantly, highlighting the effect of the metalation on the *tpy* group; the pristine *tpy* shows this repulsive feature mainly at the centre of the distal *pyr* rings (blue ticks), while the metalated *tpy-Fe* shows the strongest feature at the axial *pyr* (green tick). The *tpy* involved in the coordination node shows the same characteristics as the *tpy-Fe* (red ticks), consistent with our claim that the configurations of the node-*tpy* is identical to that of an unchained *tpy-Fe*. The exact symmetry of such tip bending effects depends strongly on the nanoscale morphology of the CO-tip apex.[5] Consequently, we only compare images acquired with the same functionalised CO tip.

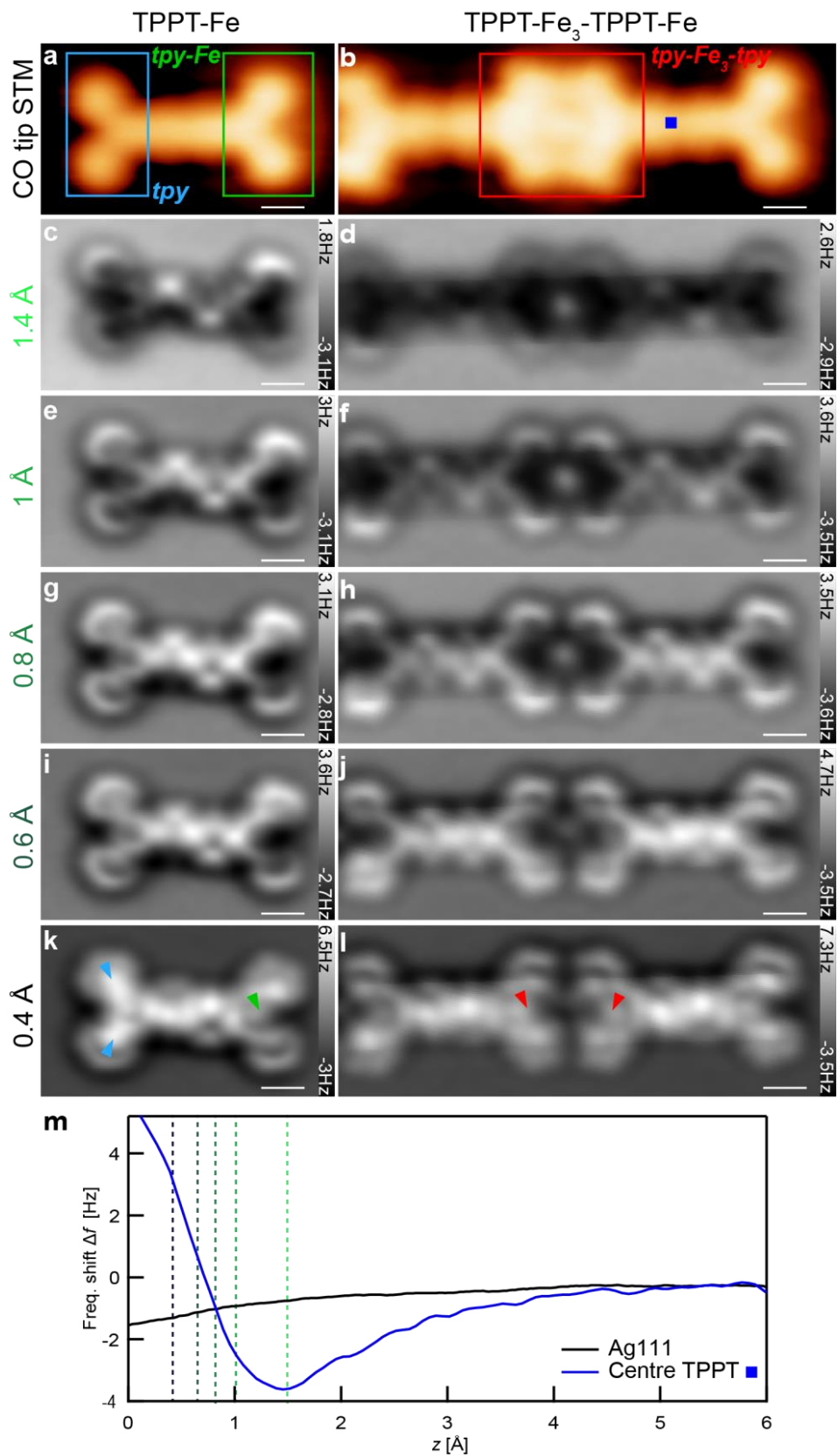

**Figure S3. CO-tip ncAFM imaging for different tip-sample distances. a,** STM topography of a singly metalated TPPT and **b,** of a coordination node in a MOC ($I_t$ = 10 pA, $V_b$ = 20 mV). **c-i,** Constant-height ncAFM images of the same systems at varying tip heights. All images were recorded with the *same* CO-functionalised Pt/Ir qPlus sensor. Gaussian filter applied. Effects due to CO-tip bending are indicated with ticks in (k) and **(**l). **m,** Frequency shift $\Delta f$ as a function of tip-sample distance $z$ measured at the centre of a TPPT molecule in a MOC [blue square in (a)] and on bare Ag(111). All data was acquired with the *same* CO-functionalised Pt/Ir tip on a qPlus sensor; $z$ = 0 refers to a tip height defined by an STM setpoint on bare Ag(111) ($I_t$ = 25 pA, $V_b$ = 20 mV). Scale bars are 5 Å.

## ncAFM image filtering process

To enhance the contrast in the ncAFM images, especially in regions with low contrast, e.g., *tpy* groups, we applied a Laplace edge detection filter based on Ref. [6]. Figure S4 displays the steps of the filtering process for the pristine TPPT, the singly metalated TPPT and the coordination node (note that these images were recorded with a resolution of 128 pixels/nm):

  i. Gaussian smoothing with full width at half maximum ~1-2 pixels (multiple times);
 ii. Second order edge detection by convolution with a Laplacian edge detection kernel -[0 ¼ 0; ¼ -1 ¼; 0 ¼ 0];
iii. Minimum filter (for sharpening), that is, selection for each pixel of a local minimum within a disk area (we used the Matlab function '*imerode*' with a disk radius of 3 pixels).

The prominent repulsive feature in the centre of the coordination node is affected by the filtering. Figure S5 shows the frequency shift $\Delta f$ measured across and along the coordination node, for each step of the filtering process. For example, the raw ncAFM data show a value of $\Delta f$ at the centre of the node, which is smaller than that for bare Ag(111) as well as that for the bonds of the adjacent, distal

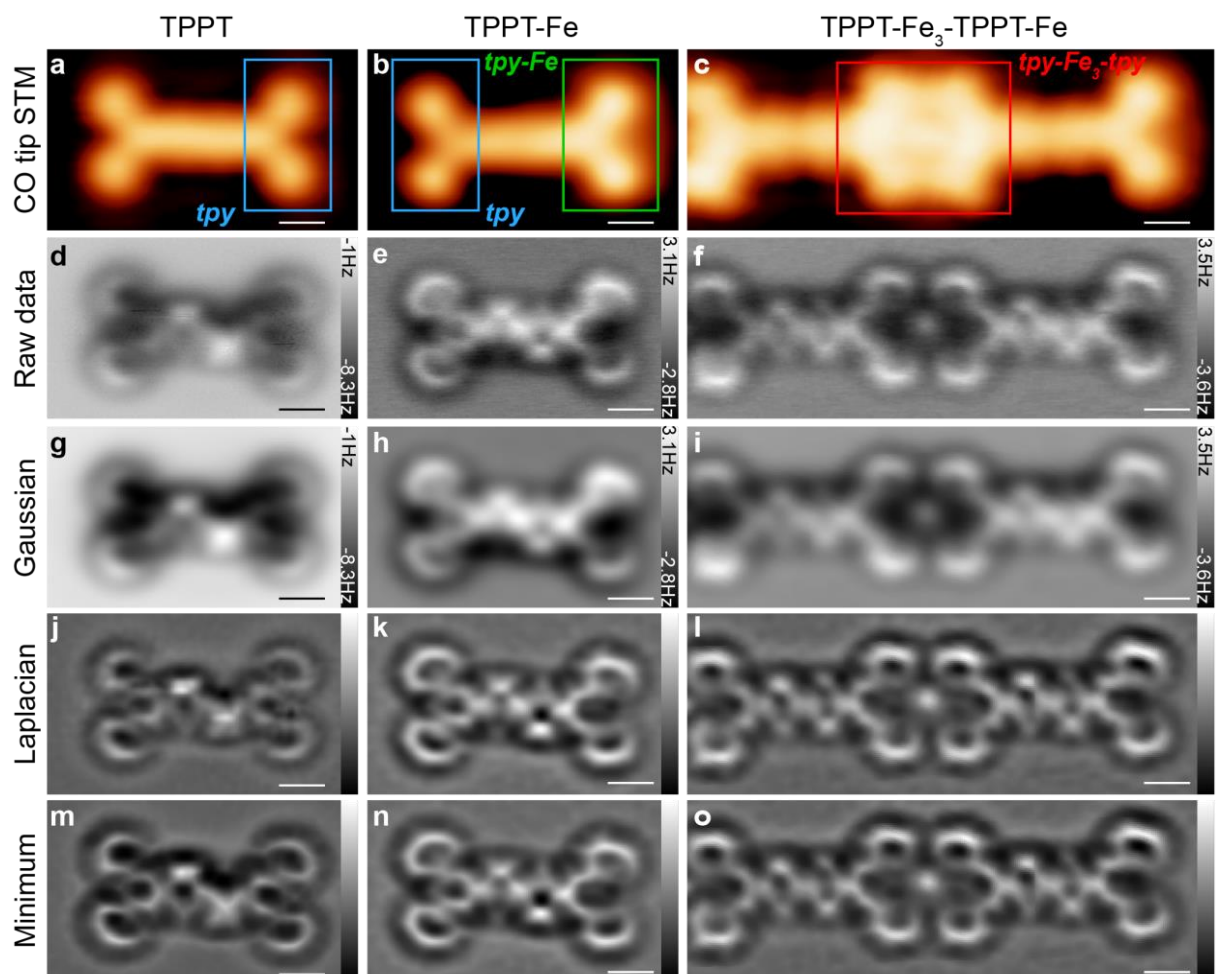

**Figure S4. NcAFM imaging filtering steps. a-c,** STM topography images with CO-functionalised Pt/Ir tip on qPlus sensor (a,b: $I_t$ = 5 pA, $V_b$ = 20 mV; c: $I_t$ = 10 pA, $V_b$ = 20 mV). **d- f,** Raw constant-height ncAFM images, 0.8 Å above an STM setpoint on bare Ag(111) ($I_t$ = 25 pA, $V_b$ = 20 mV). **g-i,** Same images with Gaussian smoothing. **j-l,** Second order Laplacian edge detection. **m-o,** Sharpening using a minimum filter. Scale bars are 5 Å (64 pixels).

*pyr*'s. The Laplacian filter, which is sensitive to the local variation of Δ*f*, enhances contrast within the node and renders the apparent height of the central feature larger the Ag(111) level. In general, the Laplace-filtered images reproduce the topographic features and increase the contrast by emphasizing the edges. It is important to note that this filtering process does not alter our interpretation of the data.

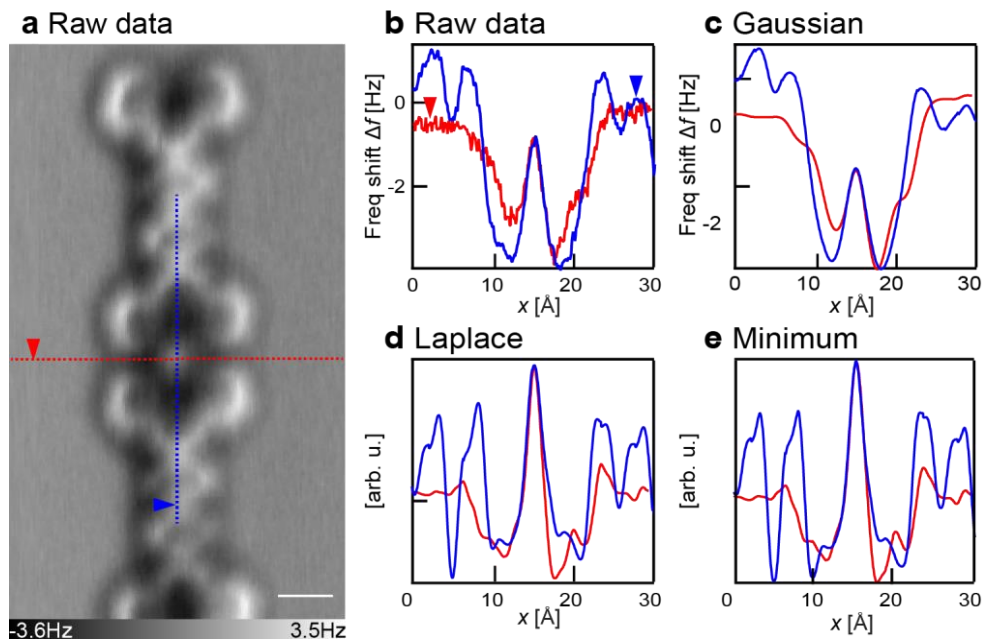

**Figure S5. Influence of the ncAFM data filtering on the imaging of the coordination node**. **a,** Raw constant-height ncAFM data with a CO-functionalised Pt/Ir tip on a qPlus sensor, 0.8 Å above an STM setpoint on the bare Ag(111) ($I_t$ = 25 pA, $V_b$ = 20 mV). **b-e,** NcAFM apparent height profiles across (red) and along (blue) the node, for different steps of the filtering process. Scale bar is 5 Å (64 pixels).

## *Adsorption geometry from ncAFM imaging*

The ncAFM maps of the TPPT reveal a non-planar adsorption geometry of the molecule (Fig. S6), an observation that is confirmed by our DFT calculations. Several molecular moieties are affected: the central phenyl (*ph*) rings (red, α) and the distal *pyr* rings (blue, β) are rotated out of the molecular plane, while the distal *pyr* change their in-plane angle as well (green, γ). To gauge the impact of metalation and chain formation on the molecular conformation, we estimate these angles from the ncAFM data and compare them to the relaxed DFT geometry, for TPPT in various stages of metalation.

We estimate the out-of-molecular-plane angles within the assumption that a given frequency shift over a C-C bond corresponds to a determined height of the tip over this bond. Based on the ncAFM maps recorded at different tip heights above the molecule (Fig. S3), we determine the apparent height difference between various points on the moiety (in Fig. S6, red arrows on *ph*, and blue arrows on *pyr*), which can then be fitted to a rotated plane. The finite height resolution of our ncAFM maps (10 pm) contributes to the measurement error. Note that the estimation of the *pyr* rotation is more challenging, since the interaction with the tip is different for nitrogen than for carbon, resulting in an uneven *pyr* plane. To address this, we determined the *pyr* out-of-plane angle by fitting a plane defined by four points on the *pyr* ring away from the nitrogen (blue and cyan arrows Fig. S6).

The in-plane angle γ of the distal *pyr*'s was directly extracted from the ncAFM maps using three anchor points: the C-C bond between *ph* and the axial *pyr* (repulsive maximum indicated by green arrow in Fig. S6b); the opposing C atom on each of the distal *pyr*'s (kink in the distal *pyr*; green arrows in Fig. S6b).

| | ncAFM | | | DFT | | |
|---|---|---|---|---|---|---|
| **TPPT** | pristine | metalated | chain | pristine | metalated | chain |
| **phenyl rotation (α)** | 3 ± 3° | 5 ± 2° | 2 ± 2° | 2.5° | 1.9° | 1.2° |
| **pyridine out-of-plane (β)** | 3 ± 2° | 4 ± 2° | 6 ± 2° | 1.5° | 9.1° | 3.3° 4.3° |
| **pyridine in-plane (γ)** | 87 ± 8° | 73 ± 6° | 80 ± 9° | 95.6° | 72° | 74° |

**Table S1. Molecular conformation changes upon adsorption, metalation, and chain formation.** Angles defined as in Fig. S6.

Table S1 shows the extracted angles for TPPT in different stages of metalation: pristine TPPT, TPPT metalated on one side and for TPPT in a MOC. We found that these experimental values are in very good agreement with the DFT geometries, for all systems. The out-of-plane rotation of the aromatic moieties is slightly underrepresented in DFT due to an overestimation of the delocalization of π electrons, leading to flatter aromatic systems.[7] Note that the DFT out-of-plane angles (β) for the distal *pyr*'s in the coordination node are slightly asymmetric for the left and right *tpy*. The reduction of the in-plane angle (γ) of the distal *pyr* by 14 ± 10° due to Fe metalation is expected and is consistent with values found in literature for other *tpy-Fe* systems.[8]

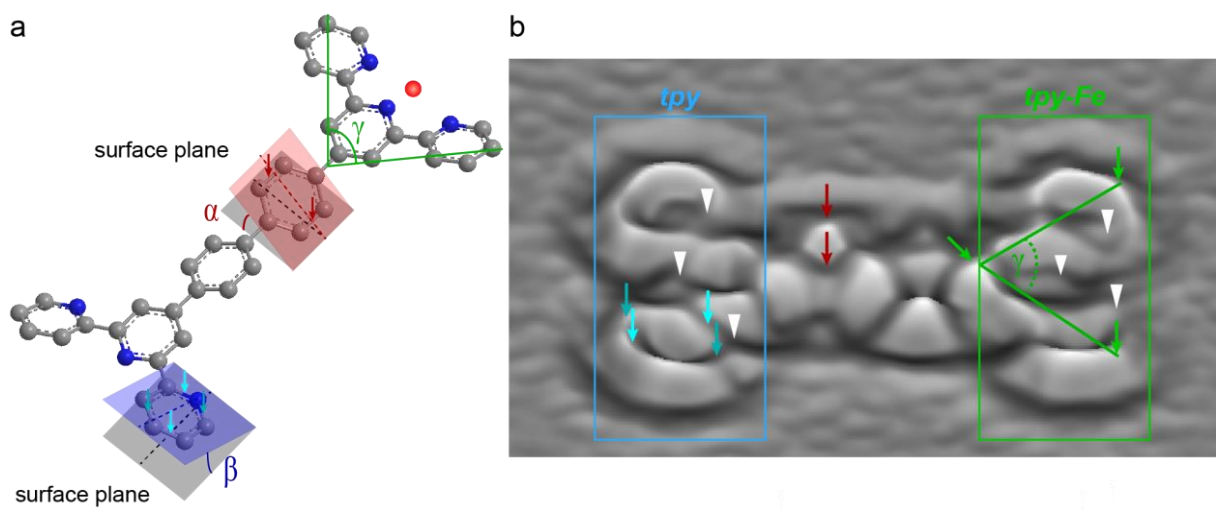

**Figure S6. TPPT adsorption geometry from ncAFM imaging. a,** Structure of singly metalated TPPT. Arrows indicate the position where height values were extracted. **b,** 3D representation of CO-tip ncAFM image of singly metalated TPPT (Fig. S4h). White ticks indicate the position of the nitrogen atoms in the *pyr* rings.

C. Krull et al.                                    Supporting Information

## Vertical ncAFM imaging

To gain insight into the height dependence of the CO-tip ncAFM central bright feature in the coordination node, we performed vertical ncAFM imaging. This approach consists of acquiring $z$-dependent frequency shift $\Delta f$ $(z)$ curves along regions of interest. The $\Delta f$ $(z)$ acquisition was cut off once $\Delta f$ reached a fixed minimum or maximum parameter, to avoid strong forces.[9,10] Vertical ncAFM imaging has been shown to resolve subtle intramolecular conformations of aromatic complexes.[29] Figure S7 shows $\Delta f$ $(z)$ curves taken at different positions $x$ along the tri-iron coordination node in Fig. 2c of the main text [$z = 0$ corresponds to an STM set point on bare Ag(111), $I_t = 5$ pA, $V_b = 20$ mV, which we estimate by $I(z)$ measurements to be $7 \pm 1$ Å above the Ag(111)]. The vertical ncAFM imaging along the node shows a shoulder (white ticks in Fig S7b), which we associate with the axial *pyr* groups of the node, indicative of the *pyr* ring bending towards the surface due to the coordination with the Fe adatoms.

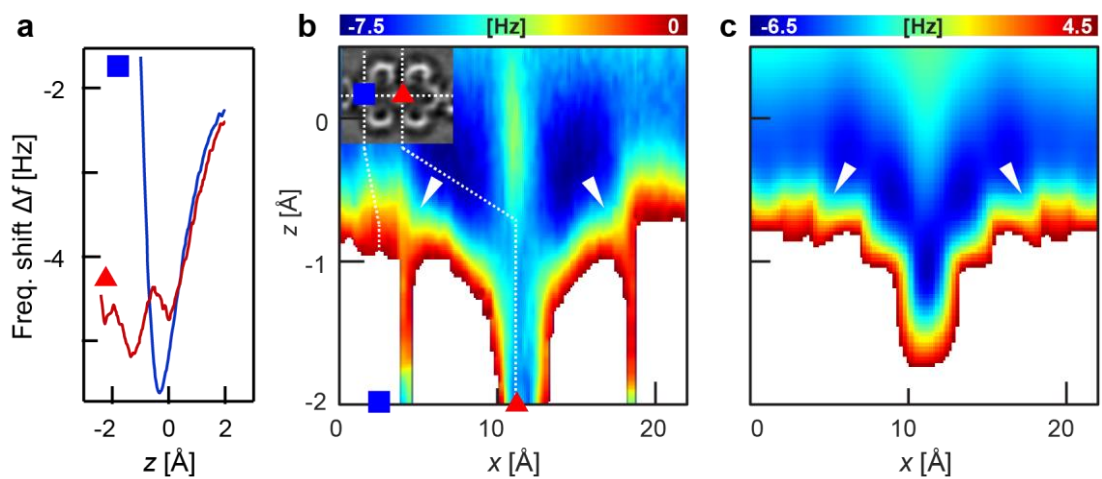

**Figure S7. Height dependence of CO-tip ncAFM along the Fe-TPPT coordination node. a,** Frequency shift $\Delta f$ as a function of tip height $z$, on TPPT in MOC (blue) and at centre of coordination node. **b,** Vertical ncAFM map across a chain node. White ticks indicate the shoulder associated with the axial metalated *pyr* group. The $z$ position is defined with respect to an STM setpoint on a bare patch of Ag(111) ($I_t = 5$ pA, $V_b = 20$ mV). NcAFM data was acquired with a CO-functionalised tip. **c,** Simulated vertical ncAFM map for a tri-iron node, reproducing the height dependence of the central feature (see Methods in main text for details on calculations).

A central repulsive feature, in between the two shoulders, is clearly visible above the molecular plane (for $z \sim 0$). This feature was not observed for *tpy*-Fe groups that are not incorporated in a chain, and is a characteristic signature of the tri-iron coordination motif. The simulated vertical frequency shift map across the coordination node exhibits the shoulders associated with the metalation of the *tpy* group (Fig. S7c, white ticks). Importantly, the simulation reproduces the height dependence of the bright central feature above the molecular plane. At this distance, electrostatic forces are the main contribution to ncAFM contrast[3], indicating an electrostatic origin of the feature.

## (d*I*/d*V*)/(*I*/*V*) STS of metalated TPPT species

Figure S8 shows (d*I*/d*V*)/(*I*/*V*) STS data for the metalated TPPT species resulting from the lateral STM manipulation in Fig. 3 of the main text. Spectra were determined by acquiring *I*(*V*) curves (with an initial STM setpoint $I_t$ = 25 pA, $V_b$ = -1 V) and calculating the numerical derivative. All data were acquired with the same Pt/Ir STM tip.

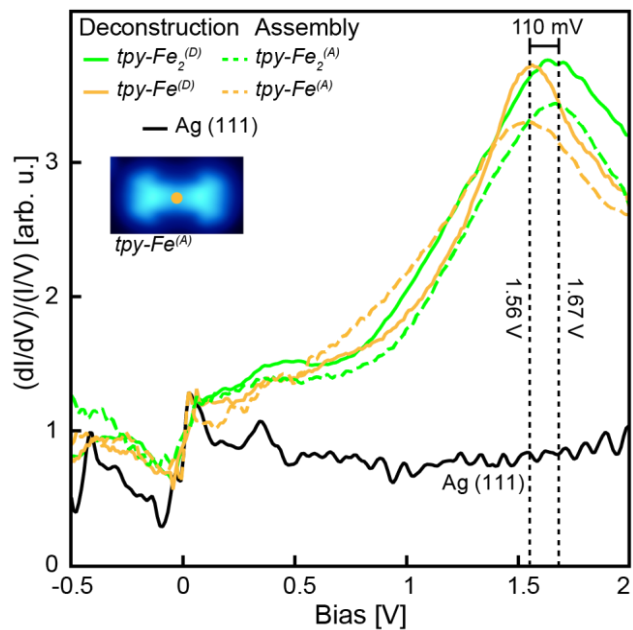

**Figure S8.** (d*I*/d*V*)/(*I*/*V*) STS spectra for assembled species *tpy-Fe*[(A)], *tpy-Fe₂*[(A)] and those resulting from the *deconstruction* of the 1D chain node via lateral STM manipulation (*tpy-Fe*[(D)], *tpy-Fe₂*[(D)]; see Fig. 3 of main text). *Tpy-Fe* (orange) consists of a *tpy* group metalated with a single Fe adatom. *Tpy-Fe₂* (green) refers to a *tpy* interacting with 2 Fe adatoms. Spectra were acquired at the centre of the TPPT ligand [see STM image of Fe-TPPT-Fe in inset ($I_t$ = 100 pA, $V_b$ = -500 mV)]. Black curve: reference Ag(111) spectrum. Data extracted from (d*I*/d*V*)/(*I*/*V*) STS in Fig. 3 of main text.

STS curves acquired at the centre of various TPPT species (inset Fig. S8) show a pronounced feature at positive biases, associated with an empty molecular orbital.[11] The TPPT molecule, which each of its *tpy* groups coordinated to a single Fe adatom (*tpy-Fe*[(A)]), shows a peak at 1.56 V. The species resulting from the deconstruction of the nanochain node (*tpy-Fe*[(D)]) shows the same spectroscopic signature; both species are equivalent. The doubly metaled *tpy-Fe₂*[(A)] moiety, assembled lateral STM manipulation (see Fig. 3 of main text) exhibits a peak at 1.67 V (110 mV larger than the singly metalated species). The *tpy-Fe₂*[(D)] species, resulting from the deconstruction of the node, shows the same feature; *tpy-Fe₂*[(A)] and *tpy-Fe₂*[(D)] are equivalent.

**Nature of the ncAFM central feature of the coordination node**

Although CO-functionalised ncAFM allows for imaging of single chemical bonds within flat aromatic molecules, imaging of single metal atoms within a cluster or metal-organic complex is more challenging. Contrast has been achieved for specific metal-organic systems (e.g., complexes on insulators,[12] metallo-tetrapyrroles[13]), but only for metal atoms in the molecular plane.[9] Usually ncAFM imaging relies on changes of molecular conformation as indirect evidence for metal-organic coordination.[14],[15],[16] The central feature observed in our CO-tip ncAFM data for the coordination node can thus – *a priori* – have a number of possible explanations:

i. Interaction of Fe with residual gas molecules. Iron centres in metal-organic complexes are known for their high affinity to small gas molecules, e.g., oxygen[17], CO[1]. However, we have identified the effect on ncAFM imaging of the specific interaction between the Fe centre of our system and CO (see Fig. S2 above). Moreover, our study was performed in UHV with an extremely low oxygen partial pressure ($p_{O2} < 1 \times 10^{-12}$ mbar), reducing the likelihood of interactions with $O_2$. We can hence rule out this hypothetical explanation.

ii. An Ag adatom from the substrate. Studies on Au(111)[16] and Cu(111)[15] have shown that reactive moieties, including *tpy*, can spontaneously form complexes with adatoms from the surface. This is not the case here, since the metalated structures only emerge after Fe deposition.

iii. A *cavity* effect. Let us assume a hypothetical scenario where there are no electrostatic interactions between the CO molecule and the tri-iron node in the MOC. That is, where the CO molecule is neutral, and tip-sample forces are of van der Waals (vdW) type. Now, let us consider the case where the ncAFM CO-tip is positioned above the node, at a height where the tip-sample vdW interactions are attractive. Since the facing *tpy*'s of the node form a *cavity* – with the central Fe lying closer to the surface than the molecular plane – the overall attractive interactions at the centre of the node can arguably be smaller than at its surrounding, resulting in an apparent saddle or protrusion, with a darker area around it (Fig. S9d). Such a feature would be enhanced by Laplace filtering. Notably, this feature would also be present in scenarios which would not involve a central Fe atom (e.g., see Fig. S10b). However, it is important to note that a central node protrusion due to this cavity effect would be a lot weaker in comparison to a protrusion given by electrostatic tip-sample interactions involving a partially charged CO (Fig. S9). The latter is in better agreement with our experiments. Moreover, a central node protrusion solely due to this cavity effect would be inconsistent with our LCPD measurements.

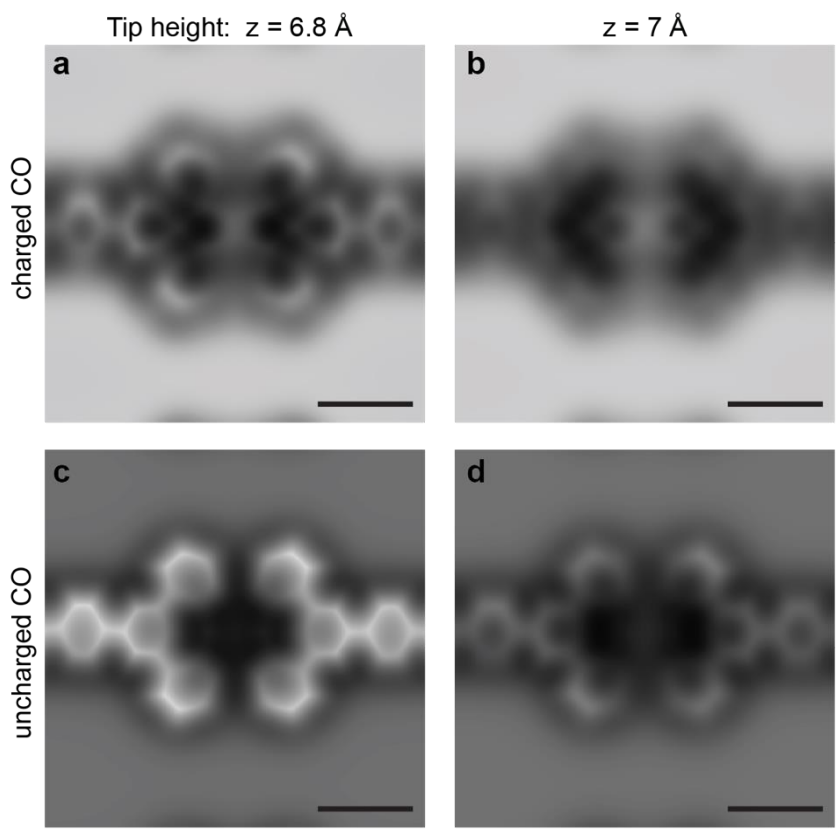

**Figure S9. Nature of the CO-tip ncAFM image protrusion at the centre of the tri-iron coordination node. a, b,** DFT-simulated constant-height CO-tip ncAFM images of the tri-iron coordination node with electrostatic tip-sample interactions given by a partially negatively charged CO molecule[18] (quadrupole charge $Q$ = -0.2 $eÅ^2$, that is, with negative lobe oriented towards sample), for different tip heights in the attractive regime. **c, d,** Same, without electrostatic forces (that is, neutral CO; $Q$ = 0). The weak central node protrusion in (d) for an uncharged CO can be explained by the cavity effect (see above); its contrast is significantly smaller than for the charged case, which is in better agreement with our experiments. The experimentally observed central node protrusion can hence be interpreted as the result of a repulsive electrostatic interaction between partially charged CO and metal-organic coordination node.

**Comparison with hypothetical di-iron coordination node**

To corroborate our claim that the metal-organic coordination node consists of a tri-iron cluster, we also performed DFT calculations for a hypothetical scenario of a di-iron coordination node. The relaxed structure of the latter involves a similar flat head-to-head coordination motif. The node consists of two opposing *tpy* groups, each metalated with a single Fe adatom, with an axial N-N distance of 8.96 Å and an Fe-Fe distance of 5.56 Å. Figure S10b shows a simulated CO-tip ncAFM image of this system. Compared to simulated ncAFM images of the tri-nuclear node (Fig. S10a), the central protrusion is less sharp, with a significantly different height profile compared to the tri-nuclear node (Fig. 2 of main text), inconsistent with our experiments.

      The calculated electrostatic potential of the di-iron node further strengthens this difference; the centre of the node exhibits a positive (that is, attractive) potential, in stark contrast to our LCPD experimental results which showed a strong negative electrostatic potential leading to a bright central feature in the CO-tip ncAFM imaging (Fig. 4 of main text). We can thus rule out the di-iron node as a possible configuration, from both the DFT calculations as well as from the STM manipulation experiments (Fig. 3 of main text).

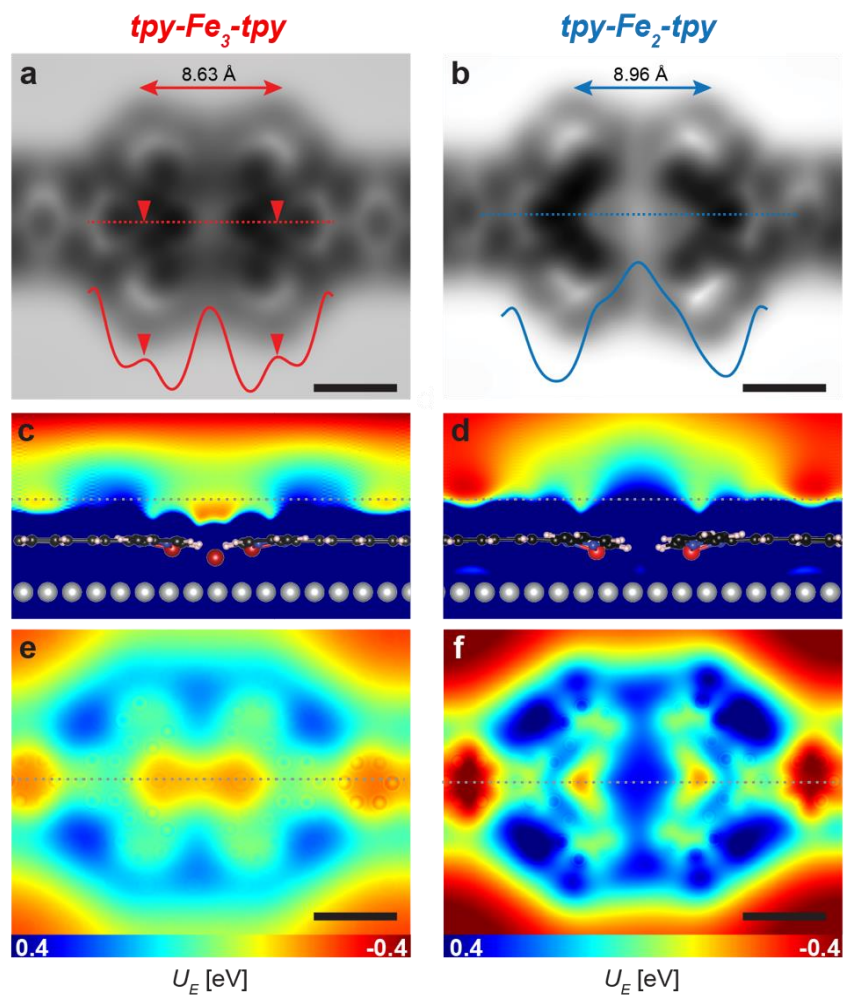

**Figure S10. Comparison between tri-iron and hypothetical di-iron coordination nodes**. **a, b,** DFT-simulated CO-tip ncAFM images of a tri-iron and di-iron node, respectively. Apparent height line profiles across the centre of the nodes and distances between the axial *pyr* N atoms are indicated. **c, d, e, f,** DFT-simulated simulated electrostatic potentials for a tri- and di-iron node, respectively (vertical and horizontal slices along dashed lines). Scale bars are 5 Å.

We would like to emphasise the importance of our multi-technique approach to fully elucidate non-trivial metal-organic structures of unknown composition. CO-functionalised ncAFM imaging alone is not sufficient to unequivocally distinguish between the di-iron and tri-iron scenarios. In this study, LCPD and STM manipulation experiments provided key data to resolve the atomic-scale structure of the MOCs.

**Bader analysis and oxidation state of iron atoms**

The concept of oxidation state of a metal atom in a metal-organic complex is useful in chemistry, materials science and condensed matter physics, since it is directly linked to its chemical reactivity. However, it is a classical concept that relies on the idea of an integer number of electrons per individual atom, and it is therefore ambiguously defined within the framework of quantum mechanics. In particular, it is not trivial to assign an oxidation state to an atom from first-principles calculations.[19] In the main text, we performed DFT calculations and assigned a Bader charge[20] to the Fe atoms depending on their position in the tri-iron coordination node (central: ~ +0.15$e$; distal: ~ +0.8$e$). Although this Bader analysis does not allow us to unambiguously determine the oxidation states of these Fe atoms, we can reliably claim that: (i) each of the three Fe atoms is positively charged; (ii) given a Bader charge difference of ~0.65$e$, the electronic configurations (and hence chemical reactivity) of distal and central Fe's differ significantly. Indeed, previous DFT-based studies on metal-organic complexes[19] were able to associate Bader charge differences on the order of ±0.2$e$ with differences in oxidation state on the order of ±1. This corroborates our claim of a mixed valence tri-nuclear coordination node.

**Van der Waals parameters for AFM simulations**

Simulated ncAFM images were obtained using the ProbeParticle code.[21] Within this model attraction due to London dispersion and Pauli repulsion are modelled by Lennard Jones potentials:

$$E_{ij}(r_{ij}) = \epsilon_{ij}\left(\left(R_{ij}/r_{ij}\right)^{12} - 2\left(R_{ij}/r_{ij}\right)^{6}\right)$$

where $\epsilon_{ij}$ and $R_{ij}$ are calculated using mixing rules from element-wise parameters (Lorentz-Berthelot):

$$\epsilon_{ij} = \sqrt{\epsilon_{ii}\epsilon_{jj}} \text{ and } R_{ij} = R_{ii} + R_{jj}.$$

The parameters used in the simulations are listed in following table:

| Element | $R_{ii}$[Å] | $\epsilon_{ii}$[meV] |
|---|---|---|
| H | 1.4870 | 0.681 |
| C | 1.9080 | 3.729 |
| N | 1.7800 | 7.370 |
| O | 1.6612 | 9.106 |
| Ag | 2.3700 | 10.000 |
| Fe | 2.0000 | 10.000 |

**Table S2. Van der Waals parameter used in the ncAFM simulations.** Note that reasonable parameters values for organic elements are well known [e.g. from Optimized Potentials for Liquid Simulations (OPLS)[22,23]] while the precise value for the metals(Ag, Fe) does not significantly impact the observed AFM contrast.